\documentclass[prd,aps,preprintnumbers,amsfonts,floatfix,nofootinbib]{revtex4}
\usepackage{epsfig}

\newcommand{\qm}{|\vec q|}
\newcommand{\qv}{{\vec q}\,}

\newcommand{\nn}{\nonumber}
\newcommand{\dd}{\displaystyle}
\newcommand{\spur}[1]{\!\not\! #1}
\begin{document}
\preprint{Napoli DSF-2004-34}

\title{\LARGE{ $ B \to D^{(\ast)}$ transitions in a quark model }}

\author{\textbf{Manuela De Vito, Pietro Santorelli}}

\affiliation{ Dipartimento di Scienze Fisiche, Universit{\`a} di
Napoli "Federico II", Italy\\
Istituto Nazionale di Fisica Nucleare, Sezione di Napoli, Italy }

\begin{abstract}
\noindent We propose a constituent quark model to evaluate heavy
decay constants and form factors relevant for $B\to D^{(\ast)}$
semileptonic transitions. We show that the model reproduces the
scaling laws dictated by the spin-flavour symmetry in the heavy
quark limit and describes quite well the experimental data.
\end{abstract}

\maketitle

\section{Introduction}
\label{s:introduction}

The study of exclusive charmed semileptonic decays of B mesons is
of primary importance to extract \cite{PDG} one of the free
parameters of the Standard Model: the absolute value of the
Cabibbo-Kobayashi-Maskawa matrix element, $|V_{cb}|$ \cite{CKM}.
The extraction is based on the prediction of the Heavy Quark
Effective Theory (HQET) \cite{isgurwise} which fix an absolute
normalization, at zero recoil point, of the form factor which
survives in the limit of infinite quark masses. Moreover, it is
possible to show that differently from $B\to D \ell \nu $ process
the $B\to D^\ast \ell \nu $ decay doesn't receive $1/m_Q$
corrections at zero recoil point \cite{LukeNeubert}. This facts
allow us to extract $|V_{cb}|$ from the differential partial decay
width for the $B\to D^{\ast} \ell \nu $ process in a nearly
model independent way \cite{CLEO,Babar}.

In this paper we will study the $B\to D^{(\ast)} \ell \nu $
processes from a different point of view. We propose a very simple
constituent quark model to evaluate heavy decay constants and
heavy-to-heavy form factors. They exhibit the scaling laws
dictated by the HQET at leading order and describe in a
satisfactory way the experimental data. To study the semileptonic
transitions between the heavy mesons $B$ and $D^{(\ast)}$ and to
compute the relevant hadronic matrix elements we use the ideas
presented in the papers in Ref. \cite{ioetalbari} devoted to study
heavy to light semileptonic and rare B transitions. In that
papers, the heavy meson B is described as a $b \bar q$  ($q\in
\{u,d\}$) bound state and the corresponding wave function,
$\psi(k)$, is obtained by solving a QCD relativistic potential
model. Here, we adopt a different point of view. As in
\cite{ioetalbari}, we describe the involved (heavy) mesons as
bound state of a heavy quark and a light anti-quark but for the
wave functions we assume their mathematical form and we fix the
free parameters by comparing model
predictions and experimental data, when available (see after).\\
The paper is organized as follows. In the next section we
introduce our constituent quark model, heavy decay constants and
heavy-to-heavy form factors are evaluated in section
\ref{s:dcfdf}. The section \ref{s:HQL} is devoted to discuss the
heavy quark limit for decay constants and form factors. Numerical
results and conclusions are collected in section \ref{s:result}.

\section{The Model}
\label{s:model}

\begin{figure}[t]
\begin{center}
\epsfig{file=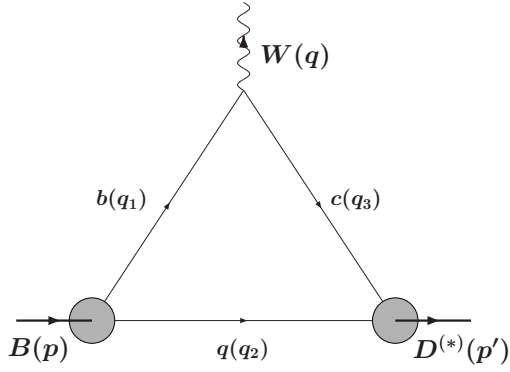,height=5cm}
\end{center}
\vspace{-0.5truecm} \caption{Quark model diagram for the
semileptonic $B$ decays involving $b\to c$ transition. The thin
lines represent quarks, the thick ones mesons. The gray disks
represent the quark-quark-meson vertexes.} \label{f:trBD}
\end{figure}

\noindent We describe any heavy meson $H(Q \overline{q})$, with
$Q\in\{b,c\}$, by introducing the matrix
\begin{equation}
H=\frac{1}{\sqrt 3}\psi_H (k){\sqrt{ \frac{m_{Q} m_{q}} {m_{q}
m_{Q} + q_1 \cdot q_2} }}\;\; \frac{\spur{q_1}+m_{Q}}{2 m_{Q}}\,
\Gamma\, \frac{-\!\!\!\spur{q_2}+m_{q}}{2 m_{q}}\,, \label{e:Hi}
\end{equation}
where $m_{Q}$ ($m_q$) stands for the heavy (light) quark mass;
$q^\mu_1,\ q^\mu_2$ their corresponding $4-$momenta (cf.
Fig.~\ref{f:trBD}). With $\psi_H(k)$ we indicate the meson's wave
function and the factors are chosen to satisfy the following
relations
\begin{eqnarray}
<H|H>& \equiv & {\rm Tr}\{(-\gamma_0 H^\dagger\gamma_0)\ H\}\ =\ 2\ m_H\, , \nonumber\\
\int \frac{d^3 k}{(2\pi)^3} |\psi_H(k)|^2& = & 2\ m_H\, .
\label{e:funnorm}
\end{eqnarray}
The meson-constituent quarks vertex, $\Gamma$, is given by
\begin{eqnarray}
\Gamma & = &  -\imath \gamma_5 \equiv \Gamma_P \hspace{5.5truecm}
\mbox{for pseudoscalar mesons} \\
\Gamma & = &
\varepsilon_\mu\left[\gamma^\mu-\frac{q_1^\mu-q_2^\mu}{m_H+m_{Q}+m_{q}}\right]
\equiv \Gamma_V(\varepsilon,q_1,q_2) \hspace{1truecm} \mbox{for
vector mesons}
\end{eqnarray}
where $\varepsilon$ is the polarization $4-$vector of the (vector)
meson $H$. In any $H Q\bar q$ vertex we assume the $4-$momentum
conservation, i.e. $q^\mu_1+q^\mu_2=p^\mu$, the H meson
$4-$momentum. Therefore, if we choose $q^\mu_1=(E_{Q},\vec k)$,
and $q^\mu_2=(E_{q},-\vec k)$, i.e. the H rest frame, we have
($k\equiv|\vec{k}|$)
\begin{equation}
E_{Q} + E_{q}\ =\ \sqrt{m_{Q}^2 + k^2} + \sqrt{m_{q}^2+ k^2}\ =\
m_H \;\; . \label{e:EnCons}
\end{equation}
which can be read as the definition of a running heavy quark mass,
as was done in Ref. \cite{ioetalbari}. In fact, the Eq.
(\ref{e:EnCons}) with the constraint $m_{Q}(k)\geq 0$ gives the
relation
\begin{equation}
0 \leq k \leq K_M\equiv \frac{m_H^2-m_q^2}{2 m_H} \label{e:kmaxI}
\end{equation}
on the loop momentum $k$
\begin{equation}
\int\frac{d^3k}{(2\pi)^3}\;\; . \label{e:loop}
\end{equation}
Let us now write down the remaining rules for the computation of
the hadronic matrix elements in the framework of this model:
\begin{itemize}
\item[a)] for the weak hadronic current, $\overline{q}_2\
\Gamma^\mu\ q_1$, one puts the factor
\begin{equation}
\sqrt{\frac{m_{q_1}}{E_{q_1}}}\ \sqrt{\frac{m_{q_2}}{E_{q_2}}}\
\Gamma^\mu\;, \label{e:J}
\end{equation}
where $\Gamma^\mu$ is some combination of Dirac matrices;
\item[b)] for each quark loop, in addition to the integration in
Eq. (\ref{e:loop}), one puts a colour factor of 3 and performs a
trace over Dirac matrices.
\end{itemize}

\section{Leptonic Decay Constants and $B\to D^{(\ast)}$ semileptonic transitions}
\label{s:dcfdf}

In this section we introduce heavy decay constants and
semileptonic form factor for heavy-to-heavy transitions
and we give their expressions in the framework of our model.\\
Using the rules introduced in the previous section we immediately
get the expressions for the heavy meson decay constants. The
pseudoscalar case was obtained and discussed in Ref.
\cite{ioetalbari}, for future convenience, we report the resulting
expression for the B meson:
\begin{equation}
f_B\ = \ \frac{\sqrt{3}}{2\pi^2 m_B^2}\int_0^{K_{M}}dk
k^2\psi_{B}(k)\frac{(m_b+m_q)(m_b m_q+q_1\cdot
q_2)}{\sqrt{E_bE_q(m_bm_q+q_1\cdot q_2)}}\, . \label{e:fB}
\end{equation}
Moreover, we have evaluate the vector heavy meson decay constant,
which is defined by
\begin{equation}
<0|V_\mu|H^\ast(p,\varepsilon)>\  =\  m_{H^\ast}\ f_{H^\ast}\
\varepsilon_\mu\,.
\end{equation}
In particular, if we consider the $B^\ast$ meson, we obtain
\begin{equation}
f_{B^\ast}\ = \ \frac{\sqrt{3}}{2\pi^2 m_{B^\ast}}\int_0^{K_{M}}dk
\frac{k^2\psi_{B^\ast}(k)}{\sqrt{E_bE_q(m_bm_q+q_1\cdot
q_2)}}\left[(m_bm_q+q_1\cdot q_2)- \frac{2}{3} \frac{k^2
m_{B^\ast}}{ m_{B^\ast}+m_b+m_q}\right]\, .
\label{e:fBst}
\end{equation}

\subsection{$B\to D$ and $B\to D^\ast$ form factors}

\noindent The same rules allow us to evaluate the matrix element
$<D(p^\prime)|\bar c \gamma_{\mu} b|B(p)>$ relevant to the weak
semileptonic transition of B to D mesons.  With reference to the
graph in Fig. \ref{f:trBD} and choosing the $4-$momenta $q_1$ and
$q_2$ as in the previous section and $q_3^\mu\ =\ (E_c, \vec k -
\qv)$, we get
\begin{eqnarray}
&& <D(p^{\prime})|\bar c \gamma_{\mu} b|B(p)> = \int_{\cal
D}\frac{d^3k}{(2\pi)^3}\psi_D(k)\psi_B(k)
\sqrt{\frac{m_q m_b}{m_q m_b+q_1\cdot q_2}}\sqrt{\frac{m_q m_c}{m_q m_c+q_3\cdot q_2}} \nn\\
&&\sqrt{\frac{m_b m_c}{E_b E_c}}\ {\rm Tr}\left[
\frac{-\spur{q_2}+m_q}{2 m_q}(\Gamma_P^\dagger) \frac{\spur{q_3}+
m_c}{2 m_c}\ \gamma_\mu\ \frac{\spur{q_1}+m_b}{2 m_b}(\Gamma_P)
\frac{-\spur{q_2}+ m_q}{2 m_q} \right ] \;\; . \label{e:BDme}
\end{eqnarray}
In the previous equation the integration domain ${\cal D}$ is
fixed enforcing the energy conservation both in the initial and
final quarks-meson vertexes. This can be done introducing, in
addition to the beauty running mass (cf. Eq. (\ref{e:EnCons})),
the charm running mass $m_c(k)$ for which $m_c(k)\geq 0$. After
some algebra the physical domain ${\cal D}$ is found to be given
by
\begin{eqnarray}
{\rm Max}(0,k_{-})\leq & k & \leq {\rm Min}(K_M,k_{+})\nonumber\\
{\rm Max}(-1,f(k,|\qv|))\leq & cos(\theta) & \leq +1 \\
0\leq & \phi & \leq 2 \pi \nonumber \label{e:domain}
\end{eqnarray}
with
\begin{eqnarray}
k_{\pm} & = & \frac{ \qm \, (m_D^2+m_q^2)\pm
(m_D^2-m_q^2) \sqrt{m_D^2+\qv^2}}{2 m_D^2}\, ,\\
f(k,\qm) & = & \frac{2 \sqrt{m_D^2+\qv^2}
\sqrt{k^2+m_q^2}-(m_D^2+m_q^2)}{2k\,\qm}\, .
\end{eqnarray}
$\phi$  and $\theta$ are the azimuthal and the polar angles,
respectively. Note that we have chosen the $z-$axis along the
direction of $\qv$, the (tri-)momentum of the W boson (cf. Fig.
\ref{f:trBD}).

The Eq. (\ref{e:BDme}) allows us to immediately extract the form
factors $f_{\pm}(q^2)$ defined by
\begin{eqnarray}
<D(p^\prime)|\bar c \gamma_{\mu} b|B(p)> & = &
f_{+}(q^2)(p_\mu+p_\mu^\prime) + f_{-}(q^2)(p_\mu-p_\mu^\prime)\,.
\label{e:F1F0}
\end{eqnarray}

The last matrix element relevant to charmed semileptonic decay of
B mesons is usually written in terms of the following form factors
\begin{eqnarray}
<D^\ast(p^\prime,\varepsilon)|\bar c \gamma_{\mu}(1-\gamma_5)
b|B(p)> & = & \!\!\!\! 2\ g(q^2)\ \epsilon^{\mu\nu\alpha\beta}\
\varepsilon^\ast_\nu\, p_\alpha\, p^\prime_\beta \nonumber \\
&-\imath & \!\!\!\! \left\{f(q^2)\, \varepsilon^\ast_\mu\ +
(\varepsilon^\ast \cdot p)\, \left
[\frac{}{}a_+(q^2)\,(p_\mu+p^\prime_\mu)+a_-(q^2)\,(p_\mu-p^\prime_\mu)\right]
\right\}\,, \label{e:gfapam}
\end{eqnarray}
they are connected in our model to
\begin{eqnarray}
&& <D^\ast(p^{\prime},\varepsilon)|\bar c \gamma_{\mu}(1-\gamma_5)
b|B(p)> = \int_{\cal
D}\frac{d^3k}{(2\pi)^3}\psi_{D^\ast}(k)\psi_B(k)
\sqrt{\frac{m_q m_b}{m_q m_b+q_1\cdot q_2}}\sqrt{\frac{m_q m_c}{m_q m_c+q_3\cdot q_2}} \nn\\
&&\sqrt{\frac{m_b m_c}{E_b E_c}}\ {\rm Tr}\left[
\frac{-\spur{q_2}+m_q}{2
m_q}(\Gamma_V(\varepsilon,q_3,q_2)^\dagger) \frac{\spur{q_3}+
m_c}{2 m_c}\ \gamma_\mu(1-\gamma_5) \frac{\spur{q_1}+m_b}{2
m_b}(\Gamma_P) \frac{-\spur{q_2}+ m_q}{2 m_q} \right ] \;\; .
\label{e:BDstme}
\end{eqnarray}
Also in this case the extraction of the form factors in Eq.
(\ref{e:gfapam}) can be done using the same frame we adopt for the
extraction of $f_\pm(q^2)$. For the polarization vectors we use:
\begin{equation}
\varepsilon^\mu(\lambda) = \left\{
\begin{array}{ll}
(0,-1,0,0) & \lambda = 1\\
(0,0,1,0) & \lambda = 2\\
(|\qv|,0,0,-E_{D^\ast})/m_{D^\ast} & \lambda = L\\
\end{array}
\right.
\end{equation}
where $E_{D^\ast} (= \sqrt{\qv^{\, 2} + m_{D^\ast}^2})$ represents
the energy of the $D^\ast$ meson.

\section{Heavy Quark Limit}
\label{s:HQL}

In this section we discuss the Heavy Quark Limit for decay
constants and form factors. We show that decay constants and
heavy-to-heavy form factors satisfy the scaling laws predicted by
HQET at leading order \cite{isgurwise}.\\
To show how the results of our model depend on the heavy quark
mass, we need to specify the shape of the wave functions
$\psi_H(k)$. We choose two possible form, the gaussian-type,
extensively used in literature (see for example
\cite{cheng,ioemisha})
\begin{equation}
\psi_H(k) = 4 \pi^{3/4}\sqrt{\frac{m_H}{\omega_H^3}}\,
\exp\left\{\frac{-k^2}{2\, \omega_H^2}\right\}\, ,
\label{e:wfgauss}
\end{equation}
and the exponential one
\begin{equation}
\psi_H(k) = 4 \pi \sqrt{\frac{m_H}{\omega_H^3}}\,
\exp\left\{\frac{-k}{\omega_H}\right\}\, , \label{e:wfexp}
\end{equation}
which is able to fit the results of relativistic quark model
\cite{salpeternardulli}. In our approach $\omega_H$ is a free
parameter which should be fixed (cf. next section for details).

\subsection{Heavy Decay Constants}

\noindent To extract the heavy mass dependence from the decay
constant, it is useful to define $x= (2 \alpha k)/m_B$ in such a
way the expressions in Eq. (\ref{e:fB}) and (\ref{e:fBst}) can be
formally written as
\begin{equation}
f_{B^{(\ast)}} = \int_0^{\alpha} dx\  \psi_B(k(x))\
F_{B^{(\ast)}}(x,z)\, , \label{e:fBlim}
\end{equation}
where $F_{B^{(\ast)}}(x,z)$ have a very simple expressions for $z
= 0$ ($z\equiv m_q/m_B$)
\begin{eqnarray}
F_B(x,0) & = &\sqrt{\frac{3}{2}}\, \frac{m_B^2}{8\pi^2\,
\alpha^3}\,
\frac{x^2(\alpha-x)}{\sqrt{(\alpha-x)(2\alpha-x)}}\,, \\
F_{B^{\ast}}(x,0) & = & \frac{m_B^2\, x^2}{8\sqrt{6}\,\pi^2\,\alpha^3}
\frac{3\left (\alpha+\sqrt{\alpha(\alpha-x)}\right)-x}
{\sqrt{2\alpha-x}\left( \sqrt{\alpha}+\sqrt{\alpha-x}\right)}
\end{eqnarray}
The integral in Eq. (\ref{e:fBlim}), for $0<\alpha\ll 1$ can be
evaluated analytically, obtaining for the leading behaviour the
following result
\begin{equation}
f_{B^{(\ast)}} \simeq \left\{
\begin{array}{lcl}
\dd \frac{1}{\sqrt{m_B}}\ \frac{\sqrt{6\, \omega_B^3}}{\pi^{3/4}} & ~~~~~~ & {\rm gaussian-type}\\
\dd \frac{1}{\sqrt{m_B}} \frac{4\sqrt{3\, \omega_B^3 }}{\pi} &
~~~~~~ & {\rm exponential-type}
\end{array}\right.
\label{e:fBHQL}
\end{equation}
in both cases in agreement with the scaling law predicted by the
HQET.

\subsection{$B\to D$ Form Factors}

The same procedure applied to the heavy-to-heavy ($B\to D$)
transitions allows us to find the scaling laws of the form factors
$f_\pm$ defined in Eq. (\ref{e:F1F0}). As for decay constants, we
can formally write
\begin{equation}
f_\pm(q^2) = \int_0^{\alpha}\; dx\; \psi_B(k(x))\; \psi_D(k(x))\
F_\pm(x,z,q^2)\, , \label{e:fpfmHQL}
\end{equation}
where, for $z=0$, $x \ll 1$ and near the zero recoil point
($q^2=q^2_{max}$)
\begin{equation}
F_\pm(x,0,q^2)|_{q^2 \simeq q^2_{max}} \simeq
\frac{x^2}{64\pi^2\alpha^3}\, m_D^2 (m_D \pm m_B) \left(
1-\frac{11}{12}(w-1) \right ).
\end{equation}
Here $w=v \cdot v^\prime$ with $v$ and $v^\prime$ the
four-velocities of the B and D mesons, respectively. Also in this
case we can extract the dependence of the form factors from the
heavy masses performing the integration in Eq. (\ref{e:fpfmHQL})
\begin{equation}
f_\pm(q^2)|_{q^2 \simeq q^2_{max}} \simeq  \frac{m_D\pm m_B}{2
\sqrt{m_D m_B}} \left\{
\begin{array}{lcl}
\dd \left [ 2\sqrt{2} \left(\frac{\omega_B\,
\omega_D}{\omega_B^2+\omega_D^2}\right)^{3/2} \left(
1-\frac{11}{12}(w-1) \right )\right ]
& ~~~~~~ & {\rm gaussian-type}\\ \\
\dd \left [ 8 \frac{\sqrt{\omega_B^3
\omega_D^3}}{(\omega_B+\omega_D)^3} \left( 1-\frac{11}{12}(w-1)
\right )\right ] & ~~~~~~ & {\rm exponential-type}
\end{array}\right.
\label{e:fpfmHQL1}
\end{equation}
It should be observed that the terms in square brackets should be
interpreted as the Isgur-Wise function, $\xi(w)$, near to $w=1$.
Moreover, in the Heavy Quark Limit we should have
$\omega_B=\omega_D$ which implies the correct normalization,
$\xi(1) = 1$, for both wave functions.

\subsection{$B\to D^\ast$ Form Factors}

\noindent The same analysis can be carried out for the $B\to
D^\ast$ form factors. Let us start to consider the form factors
$a_\pm(q^2)$. As for the $f_\pm$, we can write
\begin{equation}
a_\pm(q^2) = \int_0^{\alpha}\; dx\; \psi_B(k(x))\;
\psi_{D^\ast}(k(x))\ A_\pm(x,z,q^2)\, , \label{e:apamHQL}
\end{equation}
where, for $z=0$, $x \ll 1$ and near the zero recoil point
\begin{equation}
A_\pm(x,0,q^2)|_{q^2 \simeq q^2_{max}} \simeq
-\frac{x^2}{64\pi^2\alpha^3}\,\frac{m_{D^\ast}^2}{m_B} \left(
1-\frac{11}{12}(w-1) \right ).
\end{equation}
Analogously to the $B\to D$ case, the heavy mass dependence can be
obtained performing the integration in Eq. (\ref{e:apamHQL}),
\begin{equation}
a_\pm(q^2)|_{q^2 \simeq q^2_{max}} \simeq \mp \frac{1}{2 \sqrt{m_D
m_B}} \left\{
\begin{array}{lcl}
\dd \left [ 2\sqrt{2} \left(\frac{\omega_B\,
\omega_D}{\omega_B^2+\omega_D^2}\right)^{3/2} \left(
1-\frac{11}{12}(w-1) \right )\right ]
& ~~~~~~ & {\rm gaussian-type}\\ \\
\dd \left [ 8 \frac{\sqrt{\omega_B^3
\omega_D^3}}{(\omega_B+\omega_D)^3} \left( 1-\frac{11}{12}(w-1)
\right )\right ] & ~~~~~~ & {\rm exponential-type}
\end{array}\right.
\label{e:apamHQL1}
\end{equation}
The behaviour with heavy masses of the vectorial form factor,
$g(q^2)$, is the same of $a_-(q^2)$ in agreement with the
prediction of Heavy Quark Symmetry. For the last axial form
factor, $f(q^2)$, our model predicts
\begin{equation}
f(q^2)|_{q^2 \simeq q^2_{max}} \simeq \sqrt{m_{D^\ast} m_B}\,
(1+w) \left\{
\begin{array}{lcl}
\dd \left [ 2\sqrt{2} \left(\frac{\omega_B\,
\omega_D}{\omega_B^2+\omega_D^2}\right)^{3/2} \left(
1-\frac{11}{12}(w-1) \right ) \right ]
& ~~~~~~ & {\rm gaussian-type}\\ \\
\dd \left [ 8 \frac{\sqrt{\omega_B^3
\omega_D^3}}{(\omega_B+\omega_D)^3}\ \left( 1-\frac{11}{12}(w-1)
\right ) \right ] & ~~~~~~ & {\rm exponential-type}
\end{array}\right. .
\label{e:fHQL}
\end{equation}
Thus all the form factors satisfy the scaling laws dictated by the
HQET. Moreover, the model predicts the following Isgur-Wise
function:
\begin{equation}
\xi(w) = 1-\frac{11}{12}(w-1) + \frac{77}{96}(w-1)^2 +
o((w-1)^3)\, ,
\end{equation}
where the quadratic term, neglected in Eqs. (\ref{e:fpfmHQL1}),
(\ref{e:apamHQL1}) and (\ref{e:fHQL}), is shown. The resulting
Isgur-Wise function satisfies both the Bjorken Sum Rule
\cite{BjorkenSR}
\begin{equation}
\rho^2 \equiv -\xi^\prime(1)= \frac{11}{12} \geq \frac{3}{4}\, ,
\end{equation}
and the lower bound on the curvature \cite{OliverCurvature}:
\begin{equation}
\sigma^2 \equiv \xi^{\prime\prime}(1) = \frac{77}{48} \geq
\frac{4}{5}\rho^2\left(1+\frac{3}{4}\rho^2\right) =
\frac{99}{80}\, .
\end{equation}

\section{Numerical results and discussion}
\label{s:result}

\noindent As we have seen in section \ref{s:dcfdf}, the
heavy-to-heavy form factors can be easily extracted with the help
of the Eqs. (\ref{e:BDme}), (\ref{e:F1F0}), (\ref{e:gfapam}) and
(\ref{e:BDstme}). Nevertheless, unlike for the decay constants,
their analytical expressions are quite long, and, for the sake of
brevity, we do not report them  here.

As already discussed in the previous section, to evaluate
numerically form factors and decay constants, we must fix the
meson wave functions, $\psi_H(k)$. For the wave function we
considered two possibilities: the gaussian and the exponential
form. In both cases, for any heavy meson, H, we have one more free
parameter, $\omega_H$. In order to determine the free parameters
of the model we proceed as follows. We neglect differences between
pseudoscalar and vector mesons in the vertex function, in other
words we put $\omega_D\ =\ \omega_{D^\ast}$. Moreover,
we neglect differences between $u$ and $d$ quark masses. In such a way
the free parameters of the model are $\omega_B$, $\omega_D$ and $m_q$.
They are adjusted by fitting the experimental values of $f_D$,
BR($B\to D\ell \nu$) and the results of lattice simulation on
the ratio $f_D/f_B$. The numerical results are collected in
Tables \ref{t:datafit} and \ref{t:datafit1}.
\begin{table}[t]
\begin{center}
\begin{tabular}{|c|c|c|c|c|}
\hline
            & Exp. or Lattice  & ~our fit~ (exp.) & ~our fit (gauss.)~    \\
\hline\hline
$f_{D}/f_{B}$ & $ 1.23\pm 0.22 $ \cite{Wittig}  & ~1.03~ & ~1.05~\\
\hline
$f_{D}$ & $\, 300^{+180+80}_{-150-40}$ MeV \cite{PDG} & ~145 MeV~ & ~145 MeV~ \\
\hline
BR($B\to D\ell \nu$) & $(2.14 \pm 0.15)\, \% $  \cite{PDG} & ~2.04\, \%~ & ~1.75\, \%~ \\
\hline
$f_{B}$ &                    & ~140 MeV~ & ~139 MeV~ \\
\hline
\end{tabular}
\end{center}
\vspace{-0.5truecm} \caption{The experimental values \cite{PDG}
and Lattice result \cite{Wittig} for the decay constants used in
the fit of the free parameters of the model. For the free
parameter we assume $\omega_D=\omega_{D^\ast}$. Moreover, we use
$|V_{cb}|=(41.3\,\pm\,1.5)\times 10^{-3}$ \cite{PDG}.
\label{t:datafit}}
\end{table}
\begin{table}[t]
\begin{center}
\begin{tabular}{|c|c|c|}
\hline
~Parameter~ & ~fitted values~ (exp.) & ~fitted values (gauss.)~    \\
\hline\hline
$m_q$     & ~311 MeV~ & ~269 MeV~ \\
\hline
$\omega_B$     & ~258 MeV~ & ~421 MeV~ \\
\hline
$\omega_D$     & ~255 MeV~ & ~347 MeV~ \\
\hline
\end{tabular}
\end{center}
\vspace{-0.5truecm} \caption{The values for the free parameters of
the model in correspondence of the best fit for both wave
functions. The two sets of values are obtained for the exponential
(exp.) and gaussian vertex (gauss.). \label{t:datafit1}}
\end{table}

\begin{figure}[ht]
\begin{center}
\begin{tabular}{ccc}
\epsfig{file=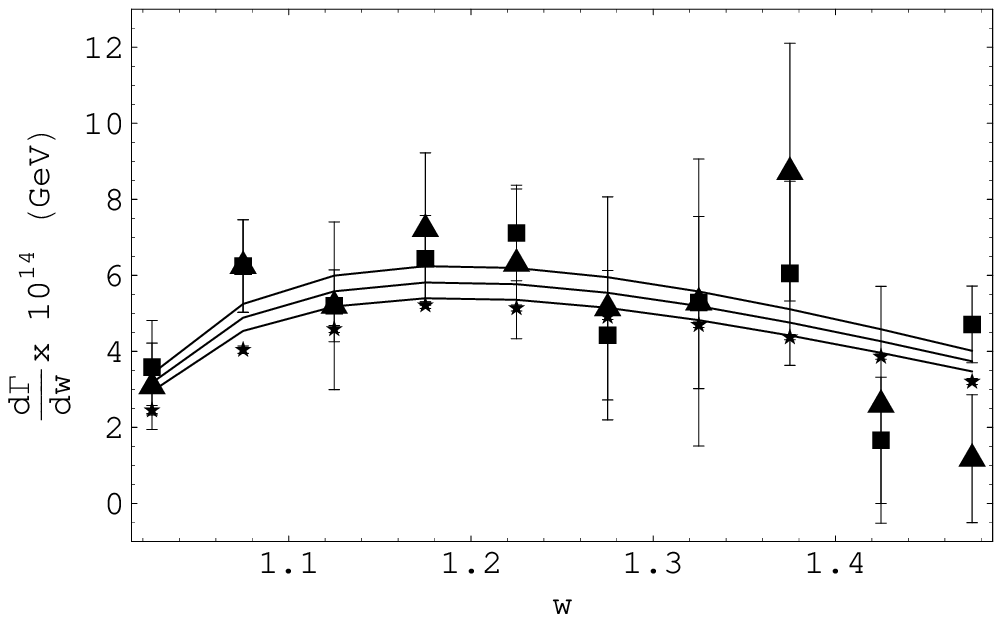,width=8truecm } & ~ &
\epsfig{file=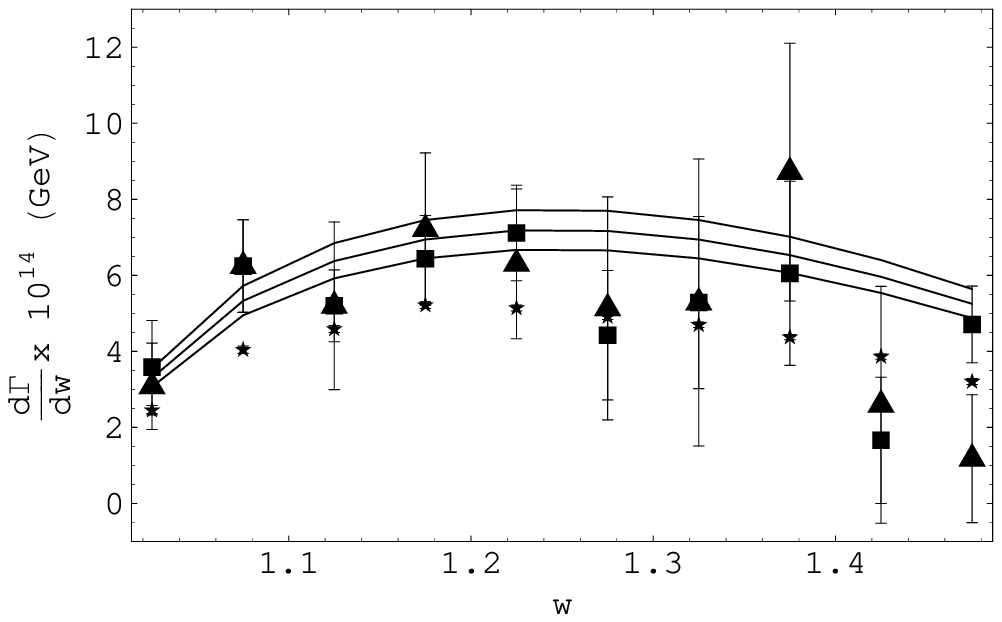,width=8truecm }
\end{tabular}
\end{center}
\vspace{-0.5truecm} \caption{Predicted ranges for $d\Gamma(B\to
D^{\ast} \ell \nu )/dw$ compared to data. Solid boxes (triangles)
refer to $\bar B^0\to D^{+\ast}\ell^-\bar\nu$ ($B^-\to
D^{0\ast}\ell^-\bar\nu$) process \cite{CLEO}. Data points from
BABAR \cite{Babar} are displayed with stars. Solid lines refer to
model predictions for exponential (left) and gaussian (right) in
correspondence of $|V_{cb}|=(39.8,41.3,42.8)\times 10^{-3}$
\cite{PDG}. \label{f:BDstar}}
\end{figure}

Comments about the results in Table \ref{t:datafit} are in order.
Let us start with decay constants. The model predicts large
$1/m_c$ corrections for $f_D$ in such a way the predicted ratio
$f_D/f_B$ violates strongly the heavy quark mass limit. However,
in the allowed region for the parameters there is the possibility
to fulfill both the heavy quark limit and the Lattice result but
the values of the decay constants ($f_D \sim f_B \sim 140$ MeV)
are predicted smaller than the ones obtained by Lattice
simulations \cite{Wittig}. Regarding the $B\to D$ form factors, it
should be observed that the experimental value for the BR($B\to
D\ell \nu$) can be reproduced with
$|V_{cb}|=(41.3\,\pm\,1.5)\times 10^{-3}$ \cite{PDG}. However,
when the exponential function is considered, the agreement becomes
better as a consequence of the larger value predicted for $f_+(0)$
($f_+(0)= 0.57$ (0.51) for exponential (gaussian) vertex
function). The same situation occurs if the differential partial
decay width for the $B\to D^{\ast} \ell \nu $ is considered. In
Figure \ref{f:BDstar}, assuming the values in Table
\ref{t:datafit1}, we plot $d\Gamma(B\to D^{\ast} \ell \nu )/dw$ in
comparison with experimental data \cite{CLEO,Babar}. In
particular, the left (right) panel allows to compare model
predictions and experimental data for the exponential (gaussian)
vertex function. Both panels contain three curves corresponding to
the predicted $d\Gamma(B\to D^{\ast} \ell \nu )/dw$ for the
central, upper and lower 1-$\sigma$ values of $|V_{cb}|$
\cite{PDG}. The agreement between model predictions and
experimental data is quite good, a better agreement requires a
smaller value of $|V_{cb}|$. In this respect, using exponential
vertex function and a value of $|V_{cb}|= 38.7\times 10^{-3}$ \cite{Babar},
the predicted $d\Gamma(B\to D^{\ast} \ell \nu )/dw$ is in very
good agreement with experimental data from Babar \cite{Babar}.

In conclusion we have proposed a constituent quark model to
describe heavy mesons. We showed that the model predictions on
decay constants and form factors reproduce the scaling laws
dictated by HQET at leading order in the heavy quark mass limit.
For finite heavy quark masses the agreement with experimental data
is quite good.

\acknowledgements{One of us (P. S.) thanks Giuseppe Nardulli for
useful discussions.}


\begin{thebibliography}{99}
%
\bibitem{PDG}
S.~Eidelman {\it et al.}  [Particle Data Group Collaboration],
Phys.\ Lett.\ B {\bf 592} (2004) 1.
%
\bibitem{CKM}
N.~Cabibbo,
Phys.\ Rev.\ Lett.\  {\bf 10} (1963) 531;
M.~Kobayashi and T.~Maskawa,
Prog.\ Theor.\ Phys.\  {\bf 49} (1973) 652.
%
\bibitem{isgurwise}
N.~Isgur and M.~B.~Wise,
Phys.\ Lett.\ B {\bf 232} (1989) 113;
N.~Isgur and M.~B.~Wise,
Phys.\ Lett.\ B {\bf 237} (1990) 527;
H.~Georgi,
Phys.\ Lett.\ B {\bf 240} (1990) 447;
A.~F.~Falk, H.~Georgi, B.~Grinstein and M.~B.~Wise,
Nucl.\ Phys.\ B {\bf 343} (1990) 1.
%
\bibitem{LukeNeubert}
M.~E.~Luke,
Phys.\ Lett.\ B {\bf 252} (1990) 447;
M.~Neubert,
Phys.\ Lett.\ B {\bf 264}, 455 (1991).
%
\bibitem{CLEO}
R.~A.~Briere {\it et al.}  [CLEO Collaboration],
Phys.\ Rev.\ Lett.\  {\bf 89} (2002) 081803
[arXiv:hep-ex/0203032].
%
\bibitem{Babar}
B.~Aubert  [BABAR Collaboration],
arXiv:hep-ex/0408027.
%
\bibitem{ioetalbari}
P.~Colangelo, F.~De Fazio, M.~Ladisa, G.~Nardulli, P.~Santorelli
and A.~Tricarico,
Eur.\ Phys.\ J.\ C {\bf 8} (1999) 81 [arXiv:hep-ph/9809372];
M.~Ladisa, G.~Nardulli and P.~Santorelli,
Phys.\ Lett.\ B {\bf 455} (1999) 283 [arXiv:hep-ph/9903206].
%
\bibitem{cheng}
H.~Y.~Cheng, C.~Y.~Cheung and C.~W.~Hwang,
Phys.\ Rev.\ D {\bf 55} (1997) 1559 [arXiv:hep-ph/9607332].
%
\bibitem{ioemisha}
M.~A.~Ivanov and P.~Santorelli,
Phys.\ Lett.\ B {\bf 456} (1999) 248 [arXiv:hep-ph/9903446];
M.~A.~Ivanov, J.~G.~Korner and P.~Santorelli,
Phys.\ Rev.\ D {\bf 63} (2001) 074010 [arXiv:hep-ph/0007169];
M.~A.~Ivanov, J.~G.~Korner and P.~Santorelli,
Phys.\ Rev.\ D {\bf 70} (2004) 014005 [arXiv:hep-ph/0311300].
%
\bibitem{salpeternardulli}
P.~Cea, P.~Colangelo, L.~Cosmai and G.~Nardulli,
Phys.\ Lett.\ B {\bf 206} (1988) 691;
P.~Colangelo, G.~Nardulli and M.~Pietroni,
Phys.\ Rev.\ D {\bf 43} (1991) 3002.
%
\bibitem{BjorkenSR}
J.D. Bjorken, invited talk at Les Rencontres de
la Vall\'{e}e d'Aoste, La Thuile, Report No. SLAC-PUB-5278, 1990;\\
N.~Uraltsev,
J.\ Phys.\ G {\bf 27} (2001) 1081 [arXiv:hep-ph/0012336];
N.~Uraltsev,
Phys.\ Lett.\ B {\bf 501} (2001) 86 [arXiv:hep-ph/0011124].
%
\bibitem{OliverCurvature}
A.~Le Yaouanc, L.~Oliver and J.~C.~Raynal,
Phys.\ Rev.\ D {\bf 69} (2004) 094022 [arXiv:hep-ph/0307197].
%
\bibitem{Wittig}
H.~Wittig,
Eur.\ Phys.\ J.\ C {\bf 33} (2004) S890
[arXiv:hep-ph/0310329].
\end{thebibliography}
\end{document}